# A novel trigon-based dual authentication protocol for enhancing security in grid environment

V. Ruckmani
Senior lecturer
Department of Computer Applications
Sri Ramakrishna Engineering College, India

Dr G Sudha Sadasivam
Professor
Department of Computer Science and Engineering
PSG College of Technology, Coimbatore, India

*Abstract*— **In recent times, a necessity has been raised in order to distribute computing applications often across grids. These applications are dependent on the services like data transfer or data portal services as well as submission of jobs. Security is of utmost importance in grid computing applications as grid resources are heterogeneous, dynamic, and multi-domain. Authentication remains as the significant security challenge in grid environment. In traditional authentication protocol a single server stores the sensitive user credentials, like username and password. When such a server is compromised, a large number of user passwords, will be exposed. Our proposed approach uses a dual authentication protocol in order to improve the authentication service in grid environment. The protocol utilizes the fundamental concepts of trigon and based on the parameters of the trigon the user authentication will be performed. In the proposed protocol, the password is interpreted and alienated into more than one unit and these units are stored in two different servers, namely, Authentication Server and Backend Server. Only when the combined authentication scheme from both the servers authenticates the user, the privilege of accessing the requested resources is obtained by the user. The main advantage of utilizing the dual authentication protocol in grid computing is that an adversary user cannot attain the access privilege by compromising a single consolidated server because of the fact that the split password is stored in different servers.**

## I. Introduction

Enhanced network bandwidth, powerful computers, and the acceptance of the Internet have motivated the constant necessity for latest and enhanced ways to compute [1]. The growing complexity of computations, improved processing power of the personal computers and the constantly rising speed of the Internet have laid down the path for grid computing [2]. "Grid" computing has emerged as a significant new field, distinguished from conventional distributed computing by its concentration on large-scale resource sharing, innovative applications, and, in some cases, high-performance orientation [3]. Grid computing is concentrating on large-scale resource sharing and collaboration over enterprises and virtual organizations boundaries [4]. As the goal of grid computing is to only provide secure grid service resources to legal users, the security issue becomes a significant concern of grid computing [5]

The necessity for secure communication between entities on the Grid has motivated the development of the Grid Security Infrastructure (GSI). GSI provides integrity, protection, confidentiality and authentication for sensitive information transferred over the network in addition to the facilities to securely traverse the distinct organizations that are part of collaboration [6]. Secure invocation of Grid services brings out the need for a security model that reflects the security components that require to be recognized and defined based on the Grid security requirements [8]. Security requirements within the Grid environment are motivated by the requirement to support scalable, dynamic, distributed virtual organizations (VOs) [3]—collections of various and distributed individuals that are looking to share and utilize different resources in a synchronized fashion [7].

A general scenario within Grid computing involves the formation of dynamic "virtual organizations" (VOs) [3] including groups of individuals as well as associated resources and services combined by a general purpose but not located inside a single administrative domain [9]. The concept of Virtual Organization (VO) [3] has been launched to define the relationships between a set of grid components comprising computing resources, data, applications and users [10]. For a VO to operate successfully participants must have control over resource sharing policies via a secure infrastructure [11]. To avoid the illegal users from visiting the grid resources, it ought to be guaranteed that strong mutual authentication is necessary for users and server [5]. Users require to know if they are interacting with the "right" piece of software or human, and that their messages will not be altered or stolen as they traverse the virtual organization (if the users have such a requirement). Users will frequently need the ability to prevent others from reading data that they have stored in the virtual organization [12].

Grid systems and applications require standard security services comprising of authentication, access control, integrity, privacy [13]. Security plays a most important role in providing the confidentiality of the communication, the integrity of data and resources, and the privacy of the user information for large scale deployment of Grid [14]. The sensitive information and resources in information systems are shielded from illegitimate access by means of the access control that is widely employed as a security mechanism [15].

At the base of any grid environment, there must be mechanisms to provide security including authentication, authorization, data encryption, and so on [33]. Authentication is the basis of security in grid [34]. Basically, authentication between two entities on remote grid nodes means that each party sets up a level of trust in the identity of the other party. In practical use, an authentication protocol sets up a secure communication channel between the authenticated parties, so that successive messages can be sent devoid of repeated authentication steps, even though it is possible to authenticate every message. The identity of an entity is typically some token or name that exclusively identifies the entity [16].

Grid technologies have adopted the use of X.509 identity certificates to support user authentication. An X.509 Certificate with its corresponding private key forms a unique credential, termed as grid credential, within the Grid. Grid credentials are utilized to authenticate both users and services [17]. In order to get the authentication from the server users and services are required to provide credentials. A credential is nothing but a piece of information that is utilized to prove the identity of a subject. Security frequently depends on the strength of the protections guarding a user's credentials. The secure storage as well as the management of these credentials is the user's responsibility. Usability, user mobility, and insufficient protection of workstations can cause major problems that often weaken the security of user credentials [18]. Passwords and certificates are some of the instances of credentials. Password-based authentication is still the most extensively used authentication mechanism, mainly due to the ease with which it can be understood by end users and implemented [19].

Password authentication is considered as one of the simplest and most convenient authentication mechanisms [22]. On the other hand, password authentication protocols are very subject to replay, password guessing and stolen-verifier attacks [20].

(1) Replay attack: A replay attack is an offensive action in which an adversary impersonates or deceives another legitimate participant via the reuse of information obtained in a protocol.

(2) Guessing attack: A guessing attack involves an adversary simply (randomly or systematically) trying passwords, one at a time, in hope that the correct password is set up. Ensuring passwords selected from an adequately large space can resist exhaustive password searches. However, the majority of the users choose passwords from a small subset of the full password space. Such weak passwords with low entropy are easily guessed by means of the so-called dictionary attack.

(3) Stolen-verifier attack: In the majority of the applications, the server stores verifiers of users' passwords (e.g., hashed passwords) instead of the clear text of passwords. The stolen-verifier attack means that an adversary who steals the password-verifier from the server can use it directly to masquerade as a legitimate user in a user authentication execution. [21].

Clearly untraceable on-line password guessing attacks and off-line password guessing attacks are the most significant considerations in designing a password-based authentication scheme [22]. A great part of protocols for password-based authenticated key exchange system are intended for a single server environment where all the information about legitimate users is stored in one server. For that reason, a credential weakness is existed in this approach due to the fact that the user's password is exposed if this server is ever compromised. A natural solution includes splitting the password between two or more servers which provides concrete security proofs for authentication protocol [23]. The dual-server model that includes two servers at the server side, one of which is a public server exposing itself to users and the other of which is a back-end server staying behind the scene; users contact only the public server, but the two servers work jointly to authenticate users [24].

This paper proposes a novel dual authentication protocol which utilizes dual servers for authentication to enhance the grid security. The novelty of the protocol is the usage of the fundamental concepts and basic elements of the trigon to authenticate. With these trigon parameters, the user credential is interpreted and then stored in two servers which provide solid security evidences for authentication protocol. The dual authentication protocol gives authentication to the grid user if and only if both the servers are mutually involved in the authentication mechanism. It is not possible to obtain the password by hacking a single server. Moreover, our protocol offers effective security against the attacks like replay attack, guessing attack and stolen-verifier attack as the user authentication is a combined mechanism of two servers. Also, it provides the security to the valid users as well as securing the user credentials, as an additional feature. Succinctly, the protocol provides secured environment while the grid user entered into the VO and the services access from the grid. The remaining of the paper is organized as follows: Section II deals with some of the existing research works which have been done so far and Section III is constituted by the proposed dual authentication protocol, user registration process and design of the authentication code with required illustrations and mathematical formulations. Section IV discusses about the implementation results and Section V concludes the paper.

## II. Related Works

Wei Jiea et al. [25] have proposed a scalable GIS architecture for information management in a large scale Grid Virtual Organization (VO). The architecture was comprised of the VO layer, site layer and resource layer: at the resource layer, information agents and pluggable information sensors were deployed on every resource monitored. The information agent and sensor approach provided a flexible framework that facilitated particular information to be captured; at the site layer, a site information service component with caching capability aggregates and maintained up-to-date information of all the resources monitored inside an administrative domain; at the VO layer, a peer-to-peer approach was utilized to construct a virtual network of site information services for information discovery and query in a large scale Grid VO. In addition to that, they proposed a security framework for the GIS, which provided security policies for authentication and authorization control of the GIS at both the site and the VO layers. Their GIS has been implemented based on the Globus Toolkit 4 as Web services compliant to Web Services Resource Framework (WSRF) specifications. The experimental results showed that the GIS presented satisfactory scalability in maintaining information for large scale Grids.

Haibo Chena et al. [26] have presented the work of Daonity which was their effort to strengthening grid security. They identified that a security service which they named behavior conformity be desirable for grid computing. Behavior conformity for grid computing was an assurance that ad hoc related principals (users, platforms or instruments) forming a grid VO should each act in conformity with the rules for the VO constitution. They applied trusted computing technologies in order to attain two levels of virtualization: resource virtualization and platform virtualization. The former was about behavior conformity in a grid VO and the latter, that in an operating system. With those two levels of virtualization working together it was possible to construct a grid of truly unbounded scale by VO together with servers from commercial organizations.

Yuri Demchenko [27] has provided insight into one of the key concepts of Open Grid Services Architecture (OGSA) and Virtual Organizations (VO). They have analyzed problems related to Identity management in VOs and their possible solution on the basis of utilizing WS-Federation and related WS-Security standards. The paper provided basic information about OGSA, OGSA Security Architecture and analyses VO security services. A detailed description was provided for WS-Federation Federated Identity Model and operation of basic services for instance Security Token Service or Identity Provider, Attribute and Pseudonym services for typical usage scenarios.

G. Laccetti and G. Schmid [28] have introduced a sort of unified approach, an overall architectural framework for access control to grid resources, and one that adhered as much as possible to security principles. Grid security implementations were viewed in the light of the model, their main drawbacks were described, and they showed how their proposal was able to prevent them. They believed that a main strategy could be to adopt both PKI (Public Key Infrastructure) and PMI (Privilege Management Infrastructure) infrastructures at the grid layer, ensured that a sufficient transfer of authentication and authorization made between the Virtual Organization and Resource Provider layers. That can be attained by expanding those features at the OS layer as system applications and services.

Xukai Zoua et al. [29] have proposed an elegant Dual-Level Key Management (DLKM) mechanism by means of an innovative concept/construction of Access Control Polynomial (ACP) and one-way functions. The first level provided a flexible and secure group communication technology whereas the second level offered hierarchical access control. Complexity analysis and Simulation demonstrated the efficiency and effectiveness of the proposed DLKM in the computational grid as well as the data grid. An example was demonstrated.

Li Hongweia et al. [30] have proposed an identity-based authentication protocol for grid on the basis of the identity-based architecture for grid (IBAG) and corresponding encryption and signature schemes. Commonly, grid authentication frameworks were attained by means of applying the standard SSL authentication protocol (SAP). The authentication process was very complex, and therefore, the grid user was in a heavily loaded point both in computation and in communication. Being certificate-free, the authentication protocol aligned well with the demands of grid computing. By means of simulation testing, it was seen that the authentication protocol was more lightweight and effective than SAP, in particular the more lightweight user side. That contributed to the better grid scalability.

Yan Zhenga et al [31] have aimed at designing a secure and effective method for grid authentication by means of employing identity-based cryptography (IBC). Nevertheless, the most extensively accepted and applied grid authentication was on the basis of the public key infrastructure (PKI) and X.509 certificates, which made the system, have lesser processing efficiency and poor anti-attack capability. An identity-based signature (IBS) scheme was first proposed for the generation of private key during grid authentication. On the basis of the proposed IBS and the IBE schemes, the structure of a grid authentication model was given, followed by a grid authentication protocol explained in detail. According to the theoretical analysis of the model and the protocol, it could be discussed that the system has enhanced both the security and efficiency of the grid authentication when compared with the traditional PKI-based and some IBC-based models.

Hai-yan Wanga. C and Ru-chuan Wanga [32] have proposed a grid authentication mechanism, which was on the basis of combined public key (CPK) employing elliptic curve cryptography (ECC). Property analysis of the mechanism in comparison to the globus security infrastructure (GSI) authentications, showed that CPK-based grid authentication, might be applied as an optimized approach towards efficient and effective grid authentication.

Our proposed work on a novel dual authentication protocol utilizes dual servers for authentication to enhance the grid security. The novelty of the protocol is the usage of the fundamental concepts and basic elements of the trigon to authenticate.

## III. PROPOSED USER REGISTRATION PROCESS FOR TRIGON-BASED DUAL AUTHENTICATION

To achieve the dual authentication, it necessitates the user to register with the Authentication server. The procedures followed in the Authentication server and the Backend server during registration of the user is as follows.

The users have to register with the Authentication server, so that it can hold a part of the interpreted password with itself and another part in the Backend server. The block diagram illustrating the registration process of the users is depicted in the Figure 1.

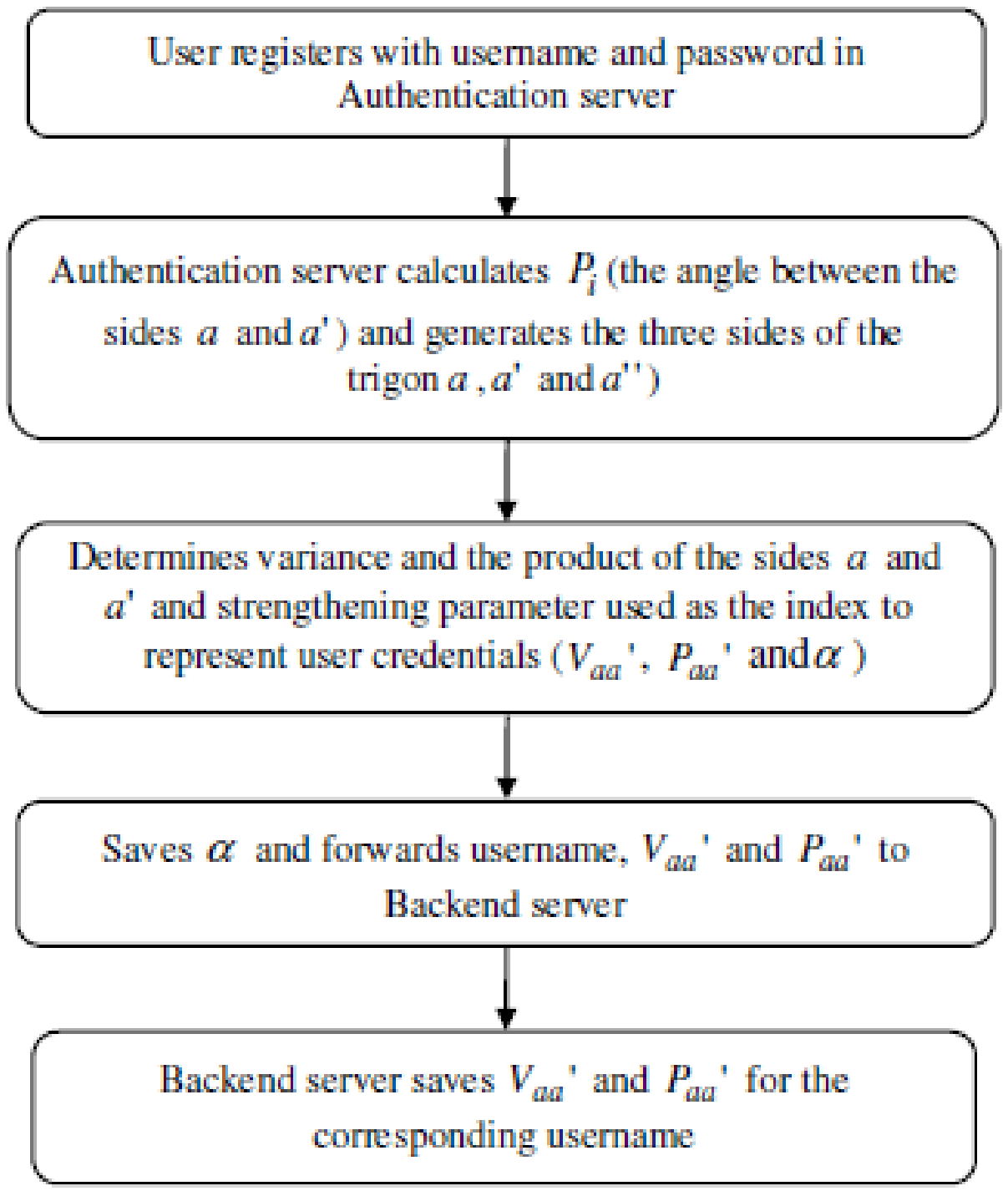

Figure 1. Flow chart explaining the registration process of user

As illustrated in Figure 1, the users who require services from the VO have to register initially with the Authentication server using their username and password. The Authentication server calculates the $P_i$ as given in (1). Along with this, the Authentication server also generates two large prime numbers, namely, $a$ and $a'$, which are considered as the two sides of a trigon. It is difficult to hack the values of $a$ and $a'$ as they are large prime numbers (as per RSA Factoring Challenge). Here, $P_i$ is taken as the angle between the two $a$ and $a'$. Now, the Authentication server can easily determine the opposite side of the angle $P_i$, termed as $a''$. With these trigon parameters, the user determines $\alpha$, $V_{aa'}$ and $P_{aa'}$ as follows

$$V_{aa'} = a - a' \qquad (1)$$

$$P_{aa'} = a * a' \qquad (2)$$

$$\alpha = 2P_{aa'} - a''^2 \qquad (3)$$

where, $a$, $a'$ and $a''$ are the three sides of the trigon,

$\alpha$ is a strengthening parameter used as the index to represent user credentials,

$V_{aa'}$ and $P_{aa'}$ are the Variance and the product of the sides $a$ and $a'$ respectively.

With the parameters $a$, $a'$ and $a''$ as the sides of trigon and $P_i$ be the angle between the sides $a$ and $a'$ the generated trigon will be assumed as in the Figure 2.

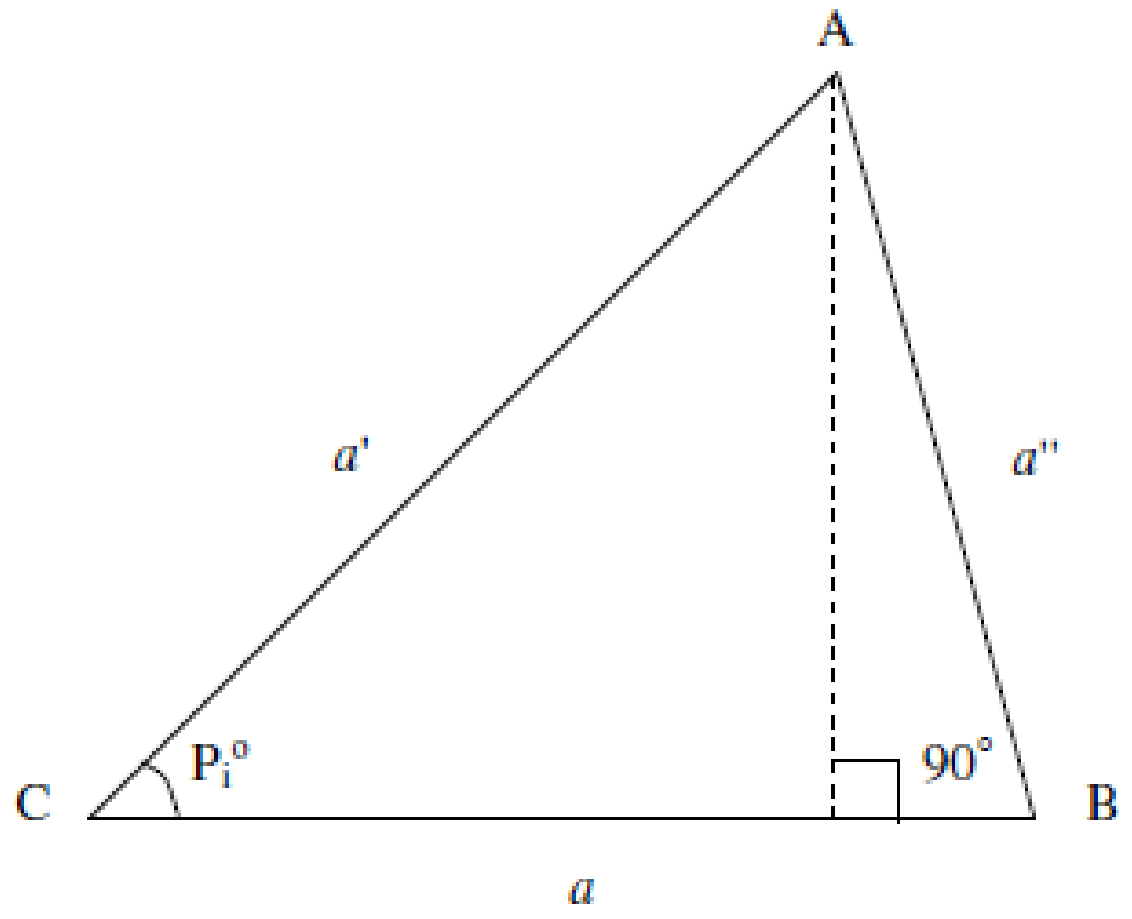

Figure 2. A sample trigon generated using the parameters $a$, $a'$, $a''$ and $P_i$

After the calculation of $\alpha$, $V_{aa'}$ and $P_{aa'}$, the authentication server stores the $\alpha$ value and its corresponding username in a database and forwards $V_{aa'}$ and $P_{aa'}$ to the Backend server along with the username. The Backend server maintains the $V_{aa'}$ and $P_{aa'}$ for the corresponding username in a database. Hence, the password is interpreted and alienated into two units and stored in two separate servers, thereby achieving the concept of dual authentication. The process is repeated for all the users who wish to register in the server, so that both the servers can maintain all the users' account. When any of the users try to access the VO, they will be validated by these servers using the account information and then allowed to access the VO by providing $T_{VO}$. If the user is an adversary and if it tries to use wrong password or username, the server can validate effectively, asserts the user as invalid and sends a warning to the adversary. The dual authentication code proposed here is designed based on the fundamentals of trigon and the design steps are discussed in the section below.

### A. *The proposed Trigon-based dual authentication protocol*

Taking the security as the main constraint in grid computing environment, we are proposing a dual authentication protocol, which will authenticate the user by a combined mechanism of two servers, namely, authentication server and backend server and then allows the user to access the VO for services. Here, the public server is mentioned as the authentication server as it performs the major authentication mechanism. The authentication procedure we have developed is on the basis of the fundamental concepts of a trigon. The Figure 3 depicts the activity diagram of the proposed dual authentication protocol.

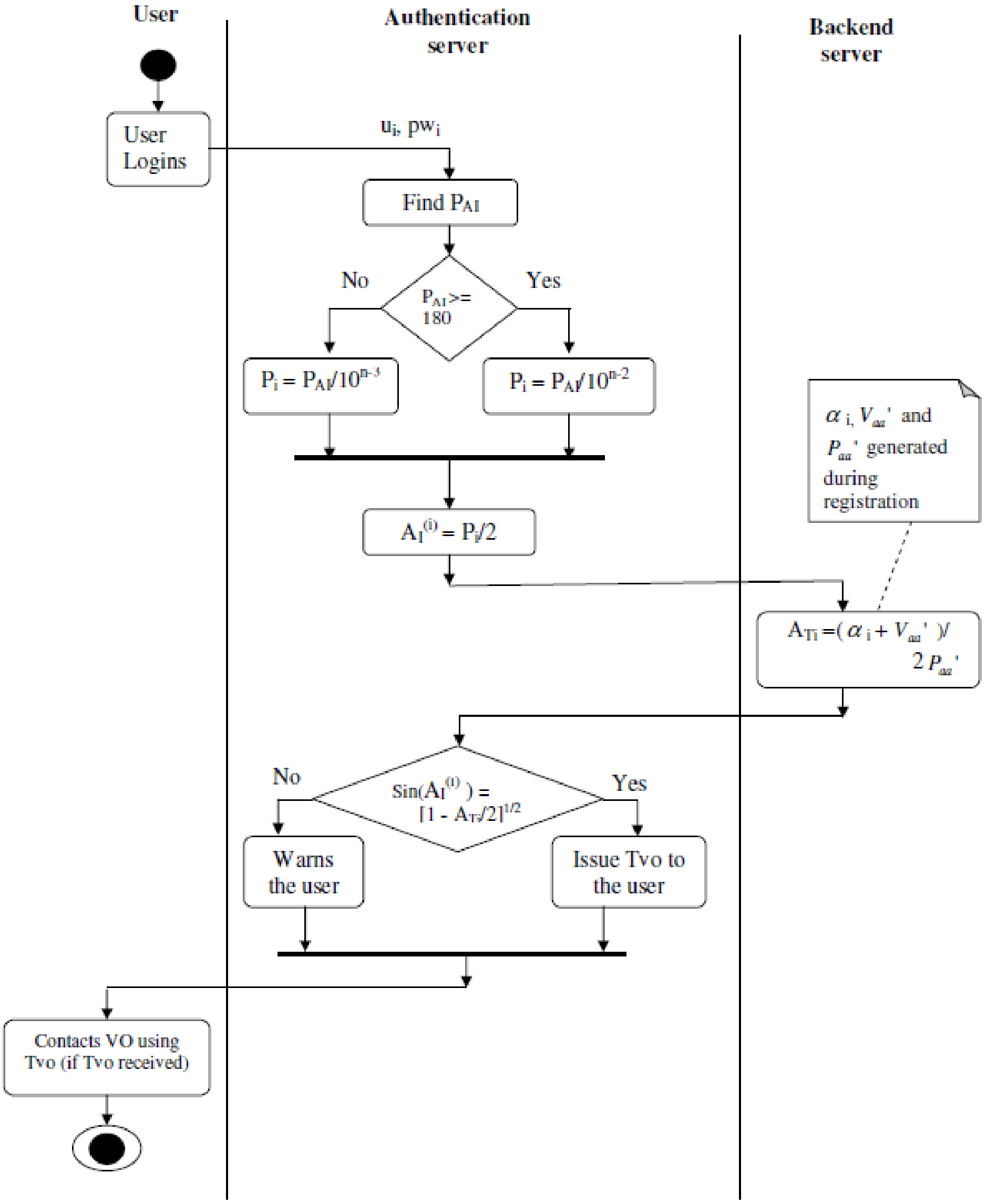

Figure 3. Activity Diagram of the proposed trigon-based dual authentication protocol

As described in the Figure 3, initially, the user who wants the services of VO has to login to the Authentication server using the username and password. Here, $u_i$ and $pw_i$ refers to username and password of $i^{th}$ user. The Authentication server calculates the Password index ( $P_i$ ) from the password as

$$P_i = \begin{cases} \dfrac{P_{AI}}{10^{n-2}}; & if \quad P_{AI}(j) \geq 180 \\ \dfrac{P_{AI}}{10^{n-3}}; & \text{else} \end{cases} \tag{4}$$

In (4), $P_{AI}$ is the ASCII-interpreted value of the given password $pw_i$, $n$ is the total number of digits in $P_{AI}$ and $P_{AI}(j)$ represents the first $j$ digits of $P_{AI}$. The $P_{AI}$ can be calculated by the following steps.

- Change the $pw_i$ into its corresponding ASCII value.
- Calculate the three-fourth of total digits of the ASCII value modulo 180, which results the first three digits of $P_{AI}$.
- Append the remaining one-fourth of the ASCII digits to $P_{AI}$.

Then, from $P_i$, the Authentication Server determines the Authentication index ( $A_I^{(i)}$ ) for $u_i$ as

$$A_I^{(i)} = \frac{P_i}{2} \tag{5}$$

Then, the Authentication Server searches for the username index $\alpha_i$ for the corresponding $u_i$ which has already been stored in the server database during the process of the registration. Subsequently, $\alpha_i$ is sent to the backend server along with $u_i$. When the Backend server receives the index $\alpha_i$ and the username from the Authentication server, it searches for $V_{aa'}$ and $P_{aa'}$ the Variance and the product of the sides $a$ and $a'$ respectively, which have been saved in the backend server database during the process of registration. From these values, the Backend server calculates an Authentication Token $A_{T_i}$ and sends it to the Authentication server to authenticate the $u_i$. The $A_{T_i}$ can be calculated as

$$A_{T_i} = \frac{\alpha_i + V^2_{aa'_i}}{2P_{aa'_i}} \tag{6}$$

In (6), $V_{aa'}$ and $P_{aa'}$ are pre-calculated values computed during individual user registration. After the retrieval of $A_{T_i}$ from the Backend server, the Authentication server authenticates the user based on the token from the Backend server and the index calculated at the Authentication server. The authentication code (or) condition which authenticates the $u_i$ is given by (proved in section III.B)

$$\mathrm{Sin}(A_I^{(i)}) = \left(\frac{1 - A_{T_i}}{2}\right)^{1/2} \tag{7}$$

The authentication process is performed by the authentication condition given in (7). When the condition is satisfied, the user is decided to be valid and the Server sends a token called Token for VO access $T_{VO}$ to the user. Using the Token $T_{VO}$ the user can contact the VO and accomplish its tasks and access the resources in the VO. If the condition is not satisfied, then a word of warning is given to the user. As a consequence, the user has no $T_{VO}$ to contact the VO and hence no resource sharing. Thus, the proposed dual authentication protocol based on two servers effectively validates the user and allows the user for resource sharing in the grid environment.

*B. Design of the authentication code*

The authentication code provided in (7) takes the eventual decision of whether the user who logins is valid or adversary. The steps by which the authentication code is developed are described elaborately as follows.

The semi-perimeter $S_P$ of the trigon depicted above is determined as

$$S_P = \frac{a + a' + a''}{2} \tag{8}$$

But it is known that,

$$Sin^2(A_I) = \left(\frac{(S_P - a) - (S_P - a')}{a \, . \, a'}\right)^{1/2} \tag{9}$$

Square of the RHS value of (9) takes the form,

$$\frac{(S_P - a) - (S_P - a')}{a \, . \, a'} = \frac{\left[\left(\frac{a + a' + a''}{2}\right) - a\right]\left[\left(\frac{a + a' + a''}{2}\right) - a'\right]}{a \, . \, a'} \tag{10}$$

Applying (9), (10) can be reorganized as follows

$$Sin^2(A_I) = \frac{2\,a\,a'}{4\,a\,a'} - \left(\frac{a^2 + a'^2 - a''^2}{4\,a\,a''}\right) \tag{11}$$

As given in (3),

$$2\,a\,a' = a''^2 + \alpha \tag{12}$$

Using (10), (9) can be written as follows,

$$Sin^2(A_I) = \frac{1}{2} - \left(\frac{a^2 + a'^2 - 2\,a\,a' + \alpha}{4\,a\,a'}\right) \tag{13}$$

$$Sin^2(A_I) = \frac{1}{2} - \left(\frac{(a - a')^2 + \alpha}{4\,a\,a'}\right) \tag{14}$$

Substituting (1) and (2), in (14) gives

$$Sin^2(A_I) = \frac{1}{2} - \left(\frac{V_{aa'}{}^2 + \alpha}{4P_{aa'}}\right) \tag{15}$$

The re-arranged format of the above equation is given by

$$Sin^2(A_I) = \frac{1}{2}\left(1 - \frac{V_{aa'}{}^2 + \alpha}{2P_{aa'}}\right) \tag{16}$$

Substituting $A_T$ which is given in (6) into (16), we can get

$$Sin^2(A_I) = \left(\frac{1 - A_T}{2}\right)^{1/2} \tag{17}$$

From the above steps, the authentication code utilized for the proposed dual authentication protocol is designed and can also serve as a proof for the effectiveness of the protocol. The protocol devised is based on the trigon parameters and effectively provides an enhanced security, because both the authentication server and the backend server have been involved in the authentication mechanism. So, compromising a single server and enjoying the VO services is impossible by any means.

## IV. Security Analysis

**Replay attack:** Usually replay attack is called as 'man in the middle' attack. Adversary stays in between the user and the server and hacks the user credentials when the user contacts server. Normally, to overcome this, the user has to change the credential randomly. But it is less probable to do that. Our protocol is robust when the replay attack happens in between the two servers as the credentials are interpreted and alienated into two parts.

**Guessing attack:** Guessing attack is nothing but the adversaries just contacts the servers by randomly guessed credentials. The effective possibility to overcome this attack is to choose the password by maximum possible characters, so that the probability of guessing the correct password can be reduced. As the proposed uses random generation of prime numbers for the calculation of the sides of the trigon, it is more difficult to guess the password.

**Stolen-verifier attack:** Instead of storing the original password, the server is normally storing the verifier of the password. If the password steals the verifier from the server, then it will masquerade as the legitimate user. But this not happens in any two server protocol, as the password is alienated into two modules. Hence, we can justify that our protocol is also more robust against the attack, as the password is interpreted and then alienated into two modules and stored in the two servers.

## V. Results and Discussions

The proposed dual authentication protocol has been implemented in the platform of JAVA (version 1.6). The protocol is tested with five valid and five invalid users. Each of the five valid users has their own username and password. Initially, they have created their user account by registering with their username and password, making them valid in the VO. The usernames, passwords and the corresponding trigon parameters of the five valid users are given in the Table II. The trigon parameters have been determined during the registration process as stated earlier and they have been stored in the database maintained at the servers.

Table I. Usernames, Passwords and the trigon parameters based on the user passwords provided at the time of registration

| Sl. No | User name | Pass word | $\alpha$ | $V_{aa'}$ | $P_{aa'}$ |
|---|---|---|---|---|---|
| 1 | user1 | admin | -3.806764915967407E11 | 665840.0 | 1.20201193169E11 |
| 2 | user2 | ascii | 2.2186644627851135E10 | 108052.0 | 1.9450880549E10 |
| 3 | user3 | test5 | 1.2148984151865493E12 | -300790.0 | 6.60523266551E11 |
| 4 | user4 | test8 | 4.213967141078015E10 | 29146.0 | 2.1752646107E10 |
| 5 | user5 | test10 | -1.9786419670597998E11 | 452092.0 | 3.377365493E9 |

The $\alpha$ values for the five valid users mentioned in the Table I have been stored in the authentication server database and $V_{aa'}$ and $P_{aa'}$ have been stored in the database of Backend server for the corresponding usernames. Instead of keeping the actual passwords, the servers maintain the interpreted passwords derived from the trigon parameters. When the servers authenticate any user, the servers determine some authentication elements based on the values which have been stored in the database and the login credential provided by the user. Using such authentication elements, the servers generate an authentication code and validate the user. The Table II shows the authentication elements generated by the servers when valid and invalid users contact the authentication server for authentication. The parameters contributed in the authentication and the authentication outcome for five valid and invalid users are given in the Table II.

TABLE II. THE AUTHENTICATION PARAMETERS DERIVED FROM THE TRIGON PARAMETERS, THE AUTHENTICATION CODE STATUS AND THE OUTCOME OBTAINED FROM AUTHENTICATION OF THE USER.

| Sl. No | User name | Pass word | $P_i$ | $A_T$ | $A_I$ | $Sin^2(A_I)$ | $\frac{1-A_T}{2}$ | Authentication code balanced? | Authentication outcome |
|---|---|---|---|---|---|---|---|---|---|
| 1 | user1 | admin | 105.11 | 0.26067301143654953 | 52.555 | 0.3696634942817 | 0.3696634942817 | Yes | Valid |
| 2 | user2 | ascii | 150.5105 | 0.8704459226549521 | 75.25525 | 0.0647770386725 | 0.0647770386725 | Yes | Valid |
| 3 | user3 | test5 | 171.1653 | 0.9881355475203079 | 85.58265 | 0.0059322262398 | 0.0059322262398 | Yes | Valid |
| 4 | user4 | test8 | 171.1656 | 0.9881363516723201 | 85.5828 | 0.0059318241638 | 0.0059318241638 | Yes | Valid |
| 5 | user5 | test10 | 164.948 | 0.965690531797593 | 82.474 | 0.0171547341012 | 0.0171547341012 | Yes | Valid |
| 6 | user1 | admins | 151.10115 | 0.26067301143654953 | 75.550575 | 0.0622628865386 | 0.3696634942817 | No | Invalid |
| 7 | user2 | asci | 59.105 | 0.8704459226549521 | 29.5525 | 0.2432668150931 | 0.0647770386725 | No | Invalid |
| 8 | user3 | test4 | 171.1652 | 0.9881355475203079 | 85.5826 | 0.0059323602682 | 0.0059322262398 | No | Invalid |
| 9 | user4 | good | 91.1 | 0.9881363516723201 | 45.55 | 0.4904012788002 | 0.0059318241638 | No | Invalid |
| 10 | user5 | user10 | 114.4948 | 0.965690531797593 | 57.2474 | 0.2926946722584 | 0.0171547341012 | No | Invalid |

The $A_I$ for each user as illustrated in the Table II, has been calculated by the authentication server and the $A_T$ for each user has been calculated by the Backend server. Based on these values, the authentication server generated the authentication code and checked whether it has been satisfied or not. When the authentication code has been satisfied by any of the user, the servers asserted that the user is valid and permits users to utilize the services offered by the VO. The status of the authentication code and the outcome of the server for valid and invalid cases are clearly tabulated in the Table II. This shows the effective performance of the protocol in enhancing the security of the grid environment by identifying valid and adversary users. Each user was provided a separate $T_{VO}$ if and only the user credential supplied by the concerned user satisfied the authentication code. The user credential that did not satisfy the authentication code was declared as invalid credential and the concerned user was asserted as an adversary. This is because that the authentication code will be satisfied if and only if the user credentials submitted for authentication are properly registered. Hence, the protocol effectively pinpointed the adversary and denied the services for that adversary user.

## VI. CONCLUSION

The authentication protocol, proposed here, enhanced the grid security as the authentication mechanism utilized two servers for authentication. As the servers kept the interpreted and distinct form of user credentials, there is very less chance to reveal the user credentials to the adversary. Moreover, the protocol utilized the fundamental properties of the trigon and the trigon parameters, made the grid more secure as the alienated passwords had been derived from these trigon parameters. This simple trigon concept utilization in the authentication protocol introduced a novel and revolutionary idea in the authentication mechanism as well as in grid environment. The implementation of our dual authentication protocol showed its effective performance in pinpointing the adversaries and paving the way to valid users for access with the VO for resource sharing. When the protocol identified any of the adversaries while authentication, it strictly prohibited those invalids from accessing with VO, which satisfies the essential pre-requisite for any authentication protocol. The development procedures of the authentication code discussed in our paper is a proof which shows the effectiveness of the protocol. So the utilization of this protocol will make the grid environment more secure.

## VII. ACKNOWLEDGEMENT

We authors would like to thank Mr. K. V. Chidambaram, Director, Data Infrastructure & Cloud Computing Group, Yahoo Software Development India Pvt Ltd., and Dr. R. Rudramoorthy, Principal, PSG College of Technology, Coimbatore for their support in carrying out the research work. We also thank the management of Sri Ramakrishna College of engineering for their support.

## REFERENCES

[1] V.Vijayakumar and R.S.D.Wahida Banu, “Security for Resource Selection in Grid Computing Based on Trust and Reputation Responsiveness”, IJCSNS International Journal of Computer Science and Network Security, Vol.8, no.11, November 2008.

[2] Wenliang Du, Jing Jia, Manish Mangal, and Mummoorthy Murugesan, “Uncheatable Grid Computing”, in Proceedings of the 24th International Conference on Distributed Computing Systems (ICDCS'04), pp. 4 - 11, 2004.

[3] Foster. I., Kesselman. C. and Tuecke. S, “The Anatomy of the Grid: Enabling Scalable Virtual Organizations”, International Journal of High Performance Computing Applications”, vol. 15, no.3, pp. 200-222, 2001.

[4] Li Wang, Wenli Wu, YingJie Li and XueLi Yu, “Content-aware Trust Statement for semantic Grid”, in proceedings of the Second International

Conference on Semantics, Knowledge and Grid, pp.95 - 95, November 2006.

[5] Rongxing Lu, Zhenfu Cao, Zhenchuan Chai, and Xiaohui Liang, "A Simple User Authentication Scheme for Grid Computing, International Journal of Network Security, vol.7, no.2, Pp.202–206, September 2008.

[6] Ionut Constandache, Daniel Olmedilla, Frank Siebenlist and Wolfgang Nejdl, "Policy-driven Negotiation for Authorization in the Semantic Grid", Technical report, L3S Research Center, October 2005.

[7] Von Welch, Frank Siebenlist, Ian Foster, John Bresnahan, Karl Czajkowski, Jarek Gawor, Carl Kesselman, Sam Meder, Laura Pearlman and Steven Tuecke, "Security for Grid Services", in proceedings of the 12th IEEE International Symposium on High Performance Distributed Computing, pp.48- 57, June 2003.

[8] N. Nagaratnam, P. Janson, J. Dayka, A. Nadalin, F. Siebenlist, V. Welch, I. Foster, S. Tuecke, The security architecture for open grid services, OGSA-SEC-WG document,http://www.cs.virginia.edu/~humphrey/ogsa-sec-wg/OGSA-SecArch-v1-07192002.pdf, July 17 2008.

[9] Foster, I., Kesselman, C., Tsudik, G. and Tuecke, S, "A Security Architecture for Computational Grids" in proceedings of the ACM Conference on Computers and Security, pp. 83-91, 1998.

[10] Thawan Kooburat and Veera Muangsin, "Centralized Grid Hosting System for Multiple Virtual Organizations", 10th Annual National Symposium on Computational Science and Engineering (ANSCSE10), Chiangmai, March 2006.

[11] David W. O Callaghan, Brian A. Coghlan, "Bridging Secure WebCom and European DataGrid Security for Multiple VOs over Multiple Grids", in proceedings of the Third International Symposium on Parallel and Distributed Computing/Third International Workshop on Algorithms, Models and Tools for Parallel Computing on Heterogeneous Networks (ISPDC/HeteroPar'04),ispdc, pp.225-231, 2004.

[12] Marty Humphrey, Mary R. Thompson and Keith R. Jackson, "Security for Grids", in Proceedings of the IEEE ,vol. 93, no. 3, pp.644 - 652, March 2005.

[13] Alexander Kemalov, "A Security Policy in GRID Architecture", International Conference on Computer Systems and Technologies, 2005.

[14] Shashi Bhanwar, and Seema Bawa, "Securing a Grid", in Proceedings of World Academy of Science, Engineering and Technology, vol.32, August 2008.

[15] M.Nithya and R.S.D.Wahida Banu, "Towards Novel And Efficient Security Architecture For Role Based Access Control In Grid Computing", IJCSNS International Journal of Computer Science and Network Security, vol. 9, no.3, March 2009.

[16] Mary R. Thompson, Doug Olson, Robert Cowles, Shawn Mullen and Mike Helm," CA-based Trust Model for Grid Authentication and Identity Delegation", Global Grid Forum CA Operations WG Community Practices Document, Oct 2002.

[17] Stephen Langella, Scott Oster, Shannon Hastings, Frank Siebenlist, Joshua Phillips,David Ervin,Justin Permar, Tahsin Kurc and Joel Saltz, "The Cancer Biomedical Informatics Grid (caBIG) Security Infrastructure", in Proceedings of 2007 AMIA Annual Symposium, Chicago, Illinois, 2007.

[18] Dr. Dennis Kafura and Dr. Markus Lorch , "A security architecture to enable user collaboration in computational grids", CISC Research Report 04-05.

[19] J. Crampton, H.W.Lim, K.G.Paterson and G.Price, "A Certificate-Free Grid Security Infrastructure Supporting Password-Based User Authentication" In Proceedings of the 6th Annual PKI R&D Workshop 2007, pp. 103-118, Gaithersburg, Maryland, USA, 2007.

[20] Lin, C.L., and T. Hwang, "A password authentication scheme with secure password updating", Computer & Security, vol.22, no.1, pp.68–72, 2003.

[21] Eun-Jun Yoon, Eun-Kyung Ryu and Kee-Young Yoo, " Attacks and Solutions of Yang et al.'s Protected Password Changing Scheme", Informatica, vol.16 , no. 2, pp. 285-294, April 2005.

[22] Her-Tyan Yeh, Hung-Min Sun and Tzonelih Hwang, "Efficient Three-Party Authentication and Key Agreement Protocols Resistant to Password Guessing Attacks", Journal of Information Science and Engineering, vol.19, no.6, pp. 1059-1070, 2003.

[23] Michael Szydlo and Burton Kaliski , "Proofs for Two-Server Password Authentication" , In proceedings of the Cryptographer's Track at the RSA(CT-RSA 2005) Conference, pp. 227-244, San Francisco, CA, USA, 2005.

[24] Yanjiang Yang, Robert H. Deng and Feng Bao, "A Practical Password-Based Two-Server Authentication and Key Exchange System", IEEE Transactions on Dependable and Secure Computing, vol. 3, no. 2, April-June 2006.

[25] Wei Jiea,Wentong Caib, Lizhe Wangc and Rob Proctera, "A secure information service for monitoring large scale grids",Parallel Computing, Vol.33, no. 7-8, pp. 572-591, August 2007.

[26] Haibo Chena, Jieyun Chenb, Wenbo Maoc and Fei Yand, "Daonity – Grid security from two levels of virtualization",Information Security Technical Report, Vol.12, no.3, pp. 123-138, 2007.

[27] Yuri Demchenko, "Virtual organisations in computer grids and identity management", Information Security Technical Report, vol.9, no. 1, pp.59-76, January-March 2004.

[28] G. Laccetti and G. Schmid, "A framework model for grid security", Future Generation Computer Systems, vol. 23, no. 5, pp.702-713,June 2007.

[29] Xukai Zoua, Yuan-Shun Dai and Xiang Rana, "Dual-Level Key Management for secure grid communication in dynamic and hierarchical groups", Future Generation Computer Systems,Vol. 23, no. 6,pp. 776-786,July 2007.

[30] Li Hongweia, Sun Shixina and Yang Haomiaoa, "Identity-based authentication protocol for grid", Journal of Systems Engineering and Electronics, Vol. 19, no. 4, pp.860-865, August 2008.

[31] Yan Zhenga, Hai-yan Wanga and Ru-chuan Wang, "Grid authentication from identity-based cryptography without random oracles", The Journal of China Universities of Posts and Telecommunications, Vol.15, no. 4, pp.55-59, December 2008.

[32] Hai-yan Wanga. C and Ru-chuan Wanga,"CPK-based grid authentication: a step forward", The Journal of China Universities of Posts and Telecommunications, Vol.14, no. 1, pp.26-31, March 2007.

[33] Yuanbo Guo, Jianfeng Ma and Yadi Wang, "An Intrusion-Resilient Authorization and Authentication Framework for Grid Computing Infrastructure",in proceedings of the Workshop on Grid Computing Security and Resource Management, Springer Berlin / Heidelberg, Vol.3516, pp.229-236, 2005.

[34] Shushan Zhao Aggarwal. A and Kent. R.D, "PKI-Based Authentication Mechanisms in Grid Systems", in proceedings of the International Conference on Networking, Architecture and Storage, pp.83-90, Guilin, July 2007.

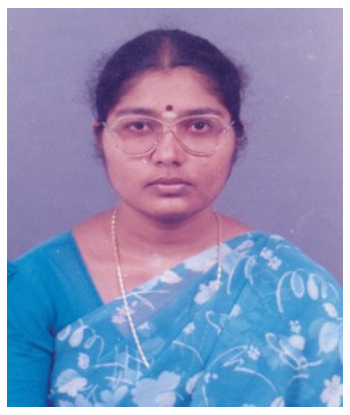

**V Ruckmani** received B. Sc, MCA and M. Phil degrees from the department of computer science, Bharathiar University, India in 1994, 1997 and 2003 respectively. She is currently pursuing the Ph. D degree, working closely with Prof. G. Sudha Sadasivam. From 1997 to 2000 she worked at PSG College of Arts and Science in the department of Computer Science. Since December 2000 she is working as a senior lecturer in Department of Computer Applications in Sri Ramakrishna Engineering College, India. She works in the field of Grid Computing specializing in the area of security. You may contact her at ruckmaniv@yahoo.com

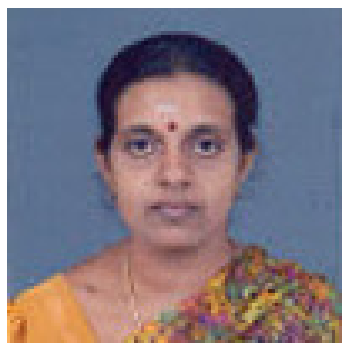

Dr **G Sudha Sadasivam** is working as a Professor in Department of Computer Science and Engineering in PSG College of Technology, India. Her areas of interest include, Distributed Systems, Distributed Object Technology, Grid and Cloud Computing. She has published 20 papers in referred journals and 32 papers in National and International Conferences. She has coordinated two AICTE – RPS projects in Distributed and Grid Computing areas. She is also the coordinator for PSG-Yahoo Research on Grid and Cloud computing. You may contact her at sudhasadhasivam@yahoo.com